\documentclass[%
 aip,
 jmp,%
 amsmath,amssymb,
preprint,%
floatfix]{revtex4-1}

\usepackage{graphicx}
\usepackage{epstopdf}
\usepackage{dcolumn}
\usepackage{bm}
\usepackage{floatrow}
\usepackage{sidecap}

\begin{document}


\title{Dispersion engineering of high-Q silicon microresonators via thermal oxidation}

\author{Wei C. Jiang}
\affiliation{
Institute of Optics, University of Rochester, Rochester, NY 14627, USA
}%
\author{Jidong Zhang}%
\affiliation{
Department of Electrical and Computer Engineering, University of Rochester, Rochester, NY 14627, USA
}%

\author{Nicholas G. Usechak}
\affiliation{
Air Force Research Laboratory, Wright-Patterson AFB, OH 45433, USA
}%

\author{Qiang Lin}%
 \email{qiang.lin@rochester.edu.}
\affiliation{
Institute of Optics, University of Rochester, Rochester, NY 14627, USA
}%
\affiliation{
Department of Electrical and Computer Engineering, University of Rochester, Rochester, NY 14627, USA
}%


\begin{abstract}
We propose and demonstrate a convenient and sensitive technique for precise engineering of group-velocity dispersion in high-Q silicon microresonators. By accurately controlling the surface-oxidation thickness of silicon microdisk resonators, we are able to precisely manage the zero-dispersion wavelength while simultaneously further improving the high optical quality of our devices, with the optical Q close to a million. The demonstrated dispersion management allows us to achieve parametric generation with precisely engineerable emission wavelengths, which shows great potential for application in integrated silicon nonlinear and quantum photonics.
\end{abstract}

\maketitle

Four wave mixing (FWM), a nonlinear parametric process mediated by the $\chi^{(3)}$ optical nonlinearity, has found many applications ranging from optical signal processing \cite{Andrekson02, AgrawalBook07} and frequency metrology \cite{Cundiff03, Dudley06}, to photonic quantum state manipulation \cite{Takesue04, Kumar05, Walmsley11, Raymer12}. In particular, FWM in high-quality (high-Q) microresonators, with dramatic cavity enhancement, enables intriguing functionalities such as ultralow-threshold parametric oscillation \cite{Vahala04, Maleki04, Moss09}, octave-spanning frequency comb generation \cite{Kippenberg07, Lipson10, Maleki11, Weiner11, Kippenberg11}, and high-purity photon-pair generation \cite{Jiang12}. However, FWM relies critically on appropriate group-velocity dispersion (GVD) to support phase matching among the interacting optical waves \cite{AgrawalBook07}. This is even more crucial in high-Q microcavities where the narrow linewidths of cavity resonances result in extremely tight tolerance for quasi-phase matching. In recent years, significant efforts have been devoted to design a variety of device structures for engineering device dispersion \cite{Knox05, Russell06, Gaeta06, Eggleton082, Osgood08, Martinelli08, Willner10, Knight10, Mas10, Gaeta11, Park12, Kippenberg12}. Nevertheless, due to the extreme sensitivity of GVD to device geometry, realization of desired dispersion in practice turns out to be fairly challenging. In this paper, we propose and demonstrate a simple but powerful approach for precise dispersion engineering in high-Q silicon microresonators. Among other applications this enables efficient parametric generation of correlated photon pairs for quantum photonic applications.

Silicon exhibits a significant Kerr nonlinearity that has attracted considerable interest in recent years to develop a variety of nonlinear functionalities \cite{Rong06, Gaeta062, Lin07, Gaeta08, Lipson08, Leuthold10, Green10, Radic10}. However, its high refractive index, although supporting tight mode confinement, leads to waveguide dispersion sensitive to device geometry, making it difficult for dispersion control due to imperfection of nanofabrication. However, one excellent property of silicon photonic devices is that their device layer thickness can be precisely controlled through thermal oxidation. This technique is widely used to produce an ideal insulating layer as a doping barrier in microelectronic devices \cite{WolfBook00}. Here we show that thermal oxidation can be employed to accurately manage the device dispersion, while simultaneously preserving or even improving the optical quality of the device.

\begin{figure}[htbp]
\includegraphics[scale=0.49]{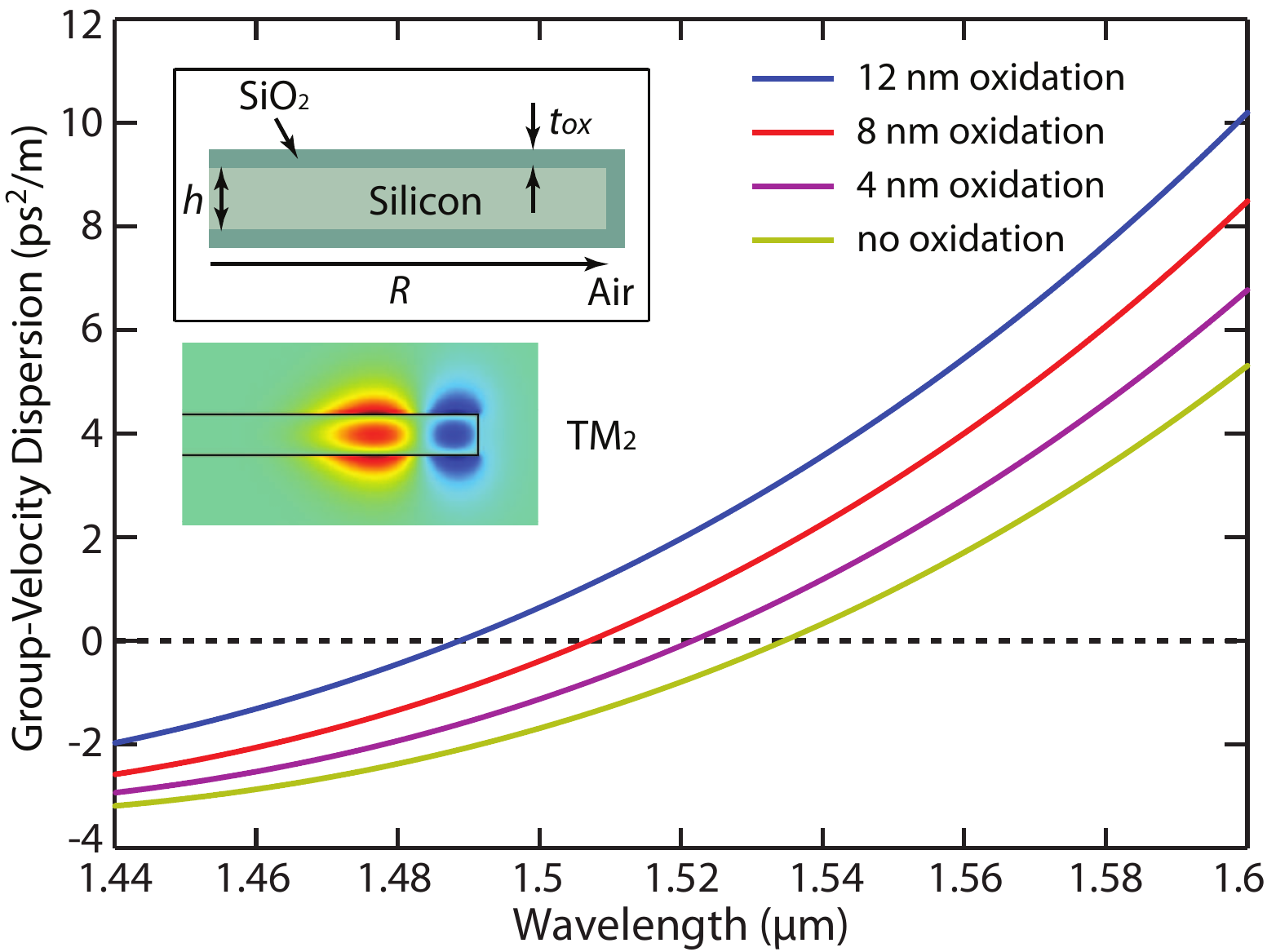}
\caption{
\label{Dispersion} Simulated GVD curves for TM${_2}$ mode of a silicon microdisk with an original thickness ${h = 260}$ nm and radius ${R = 4.5~\mu}$m for different oxidation thicknesses. Insets show the cross section of a silicon microdisk edge with a conformal thermally grown SiO${_2}$ overlayer, and simulated optical field profile for TM${_2}$ mode of the silicon microdisk.}
\end{figure}

The device structure we investigate is a silicon microdisk resonator which supports high-Q whispering-gallery optical modes. The GVD of a microdisk is dominantly determined by the disk thickness. Thermal oxidation \cite{WolfBook00} consumes silicon by a thickness of 0.44${t_o}$${_x}$ and grows a conformal SiO${_2}$ overlayer with a thickness of ${t_o}$${_x}$ covering the disk core (Fig.~\ref{Dispersion}, inset). This benefits the dispersion engineering in two ways. First, the thickness reduction of the silicon core changes the waveguide confinement, thus modifying the device dispersion. Second, the addition of the oxide overlayer covering on the surface slightly adjusts the waveguide boundary, thus offering a further dispersion modification. As the amount of thermal oxidation can be manipulated in a precise manner by controlling the oxidation time, we can thus engineer the device GVD in a very accurate fashion. For example, Fig.~\ref{Dispersion} shows the simulated GVD of the second-radial-order transverse-magnetic (TM${_2}$) mode for a silicon microdisk of different oxidation thicknesses. With an original thickness h = 260 nm and radius ${R = 4.5~\mu}$m without any oxidation, the device exhibits a ZDWL around 1535 nm. However, thermal oxidation of the device by consuming silicon layer of 1.76, 3.52, and 5.28 nm (corresponding to creating an oxide overlayer of 4, 8, and 12~nm), is able to shift the ZDWL to a shorter wavelength at 1522, 1507, and 1489~nm, respectively.

\begin{figure*}[t!]
\includegraphics[scale=0.435]{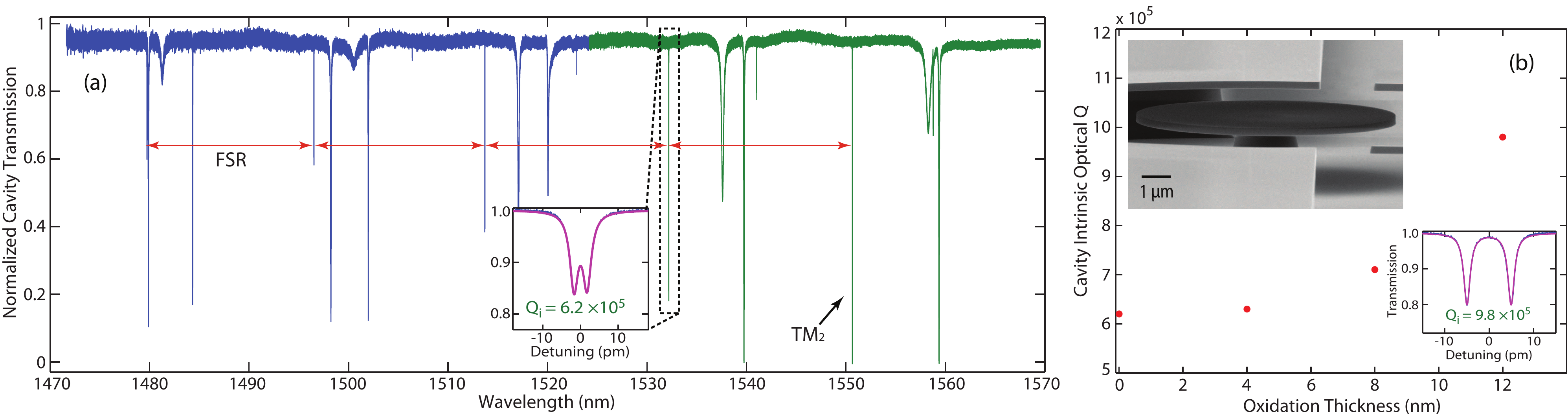}
\caption{\label{Transmission} (a) The normalized transmission spectrum of the microdisk with no oxidation scanned by two tunable lasers (indicated as blue and green) from 1470 nm to 1570 nm. The TM${_2}$ mode family is indicated by marked FSRs. The inset shows detailed transmission of a TM${_2}$ mode at 1532.2 nm with theoretical fitting in red. (b) Measured intrinsic optical Qs versus oxidation thicknesses for the TM${_2}$ mode of silicon microdisks (in red dots). The upper inset shows an SEM picture of the fabricated silicon microdisk with ${R = 4.5~\mu}$m on a silica pedestal, and the lower one shows the fitted cavity transmission with a doublet splitting of the TM${_2}$ mode for the device with oxidation thickness of 12 nm. }
\end{figure*}

To demonstrate the proposed dispersion engineering scheme, we fabricate four sets of silicon microdisk resonators with radii ${R = 4.5~\mu}$m on a standard silicon-on-insulator (SOI) wafer, with a top silicon layer thickness of 260 nm and a buried oxide thickness of 2 ${\mu}$m. The microdisk pattern is defined by the electron beam lithography with ZEP520A positive resist, and then transferred to the 260~nm silicon layer by the fluorine-based inductively-coupled-plasma (ICP) reactive-ion-etching (RIE) using C${_4}$F${_8}$/SF${_6}$ chemistry. The etching parameters are optimized to achieve a smooth device sidewall. Subsequently, the buried oxide layer is isotropically etched by using hydrofluoric (HF) acid to form a silica pedestal. A scanning electron microscope (SEM) image of a fabricated microdisk is shown in Fig.~\ref{Transmission}(b). Thermal dry oxidation of silicon is then performed separately on three sets of microdisks at 900$^{\circ}$C in the O${_2}$ ambience for 6, 15, and 26 minutes, to create a conformal oxide overlayer of thickness 4, 8, and 12 nm, respectively. The fourth set of microdisks without oxidation are used as a reference.

To characterize the optical properties of the fabricated microdisk resonators, a mode-hop-free continuous-wave tunable laser is launched into the devices by near-field evanescent coupling through a tapered optical fiber (typical diameter is about 1 ${\mu}$m), and the cavity transmission spectrum is obtained by scanning the laser wavelength between 1470 nm and 1570 nm, which is calibrated by a Mach-Zehnder interferometer. Fig.~\ref{Transmission}(a) shows the normalized optical transmission spectrum of the microdisk resonator with no oxidation. Different mode families can be identified by their free-spectral ranges (FSRs). For the TM${_2}$ mode, a high optical quality is measured consistently over the broad scanning spectral range. The inset of Fig.~\ref{Transmission}(a) shows the detailed cavity transmission of a TM${_2}$ mode at 1532.2 nm, indicating a measured intrinsic optical quality of Q$_i = 6.2\times 10^5$. Moreover, the measured Q${_i}$ increases with the oxidation thickness as shown in Fig.~\ref{Transmission}(b). For example, a higher intrinsic optical quality of $ Q_i = 9.8\times 10^5$ is achieved for the microdisk with 12~nm conformal oxide overlayer, clearly showing the advantage of the silicon thermal oxidation treatment for improving the device sidewall quality \cite{Cerrina01}.

In general, the dispersion of a microresonator can be characterized by the frequency mismatch between adjacent FSRs ${\Delta\nu=\nu _{m+1} - 2\nu _{m} + \nu _{m-1}}$ for each cavity resonance frequency ${\nu _{m}}$ with mode number ${m}$. For a micro-resonator cavity, the GVD parameter ${\beta_2}$ is closely related to ${\Delta\nu}$ given by,
\begin{eqnarray}
\beta_2 = -\frac{\Delta\nu}{\nu _{FSR}^3(2\pi)^2 R} , \label{Threshold}
\end{eqnarray}
where ${\nu _{FSR} = \nu _{m+1} - \nu _{m}}$ is the FSR of the resonator, and ${R}$ is the radius of the resonator.

%

\begin{figure}[htbp]
\includegraphics[scale=0.50]{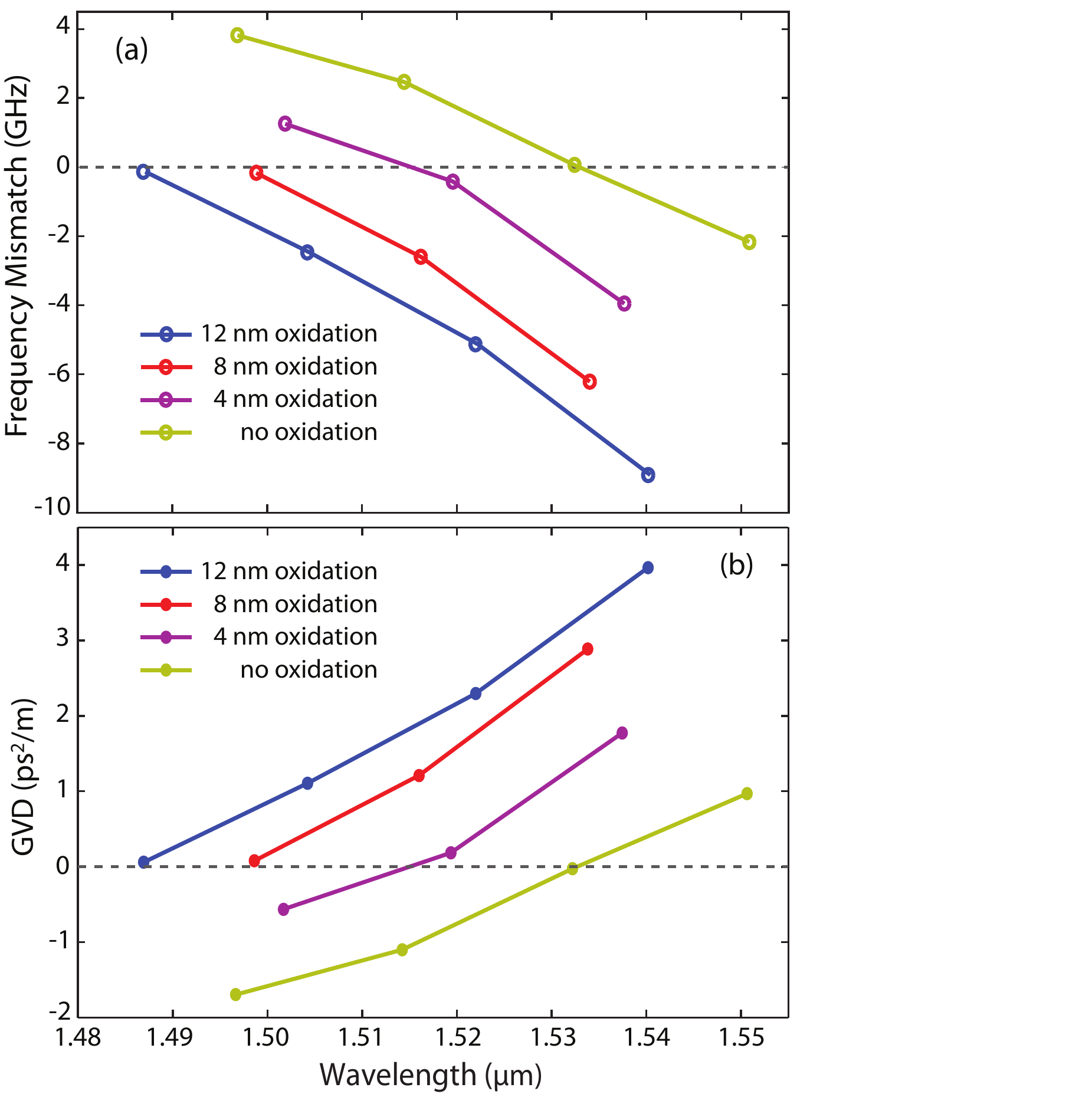}
\caption{\label{MeasuredGVD} The measured (a) frequency mismatch ${\Delta\nu}$ and (b) the corresponding GVD parameter for the TM${_2}$ mode of the microdisks with different oxidation thicknesses.}
\end{figure}

Fig.~\ref{MeasuredGVD} shows the measured ${\Delta\nu}$ and the corresponding GVD for the TM${_2}$ mode for each of the four microdisks with different oxidation conditions. It shows clearly that the GVD curve is tuned toward shorter wavelengths as oxide thickness increases. Accordingly, the ZDWL is tailored from 1532 nm (no oxidation) to 1515, 1499, and 1487 nm for thermally grown oxide thickness of 4, 8, and 12 nm, respectively. This corresponds to a ZDWL tuning rate of ${\sim3-4}$ nm per nanometer of silicon oxidation which, to the best of our knowledge, is the most accurate dispersion engineering demonstrated to date \cite{Knox05, Russell06, Gaeta06, Eggleton082, Osgood08, Martinelli08, Willner10, Knight10, Mas10, Gaeta11, Park12, Kippenberg12}. In particular, the frequency mismatch ${\Delta\nu}$ for certain cavity modes can be tuned to be within the corresponding cavity liewidth, which will ensure optimum quasi-phase matching for FWM. For example, an oxidation thickness of 8 nm is able to achieve a $\Delta \nu$ of -0.17 GHz around 1498.6 nm for the third set of devices, which is much smaller than the cavity linewidth of ${\sim0.5}$ GHz. In practice, since silicon thermal oxidation is able to provide nanometer-scale thickness control on the device, it would enable very precise dispersion control, \emph{e.g.}, to accurately compensate the frequency mismatch at the desired cavity resonance induced by fabrication imperfection.

\begin{figure*}[t!]

{\caption{\label{PLSetup} (a) Experimental setup for measuring spontaneous FWM. The CWDM MUX/DEMUX has a 3-dB bandwidth of 17 nm for each of its transmission bands whose center wavelengths are separated by 20 nm apart with a band isolation $>120$ dB. (b) The PL spectrum when pumping at 1532.2 nm with input pump power of 45.6 ${\mu}$W for the microdisk with no oxidation and (c) the PL spectrum when pumping at 1486.9 nm with input pump power of 36.2 ${\mu}$W for the device with oxidation thickness of 12 nm. For both cases, the optical pump is critically coupled to the cavity.}}
{\includegraphics[scale=0.52]{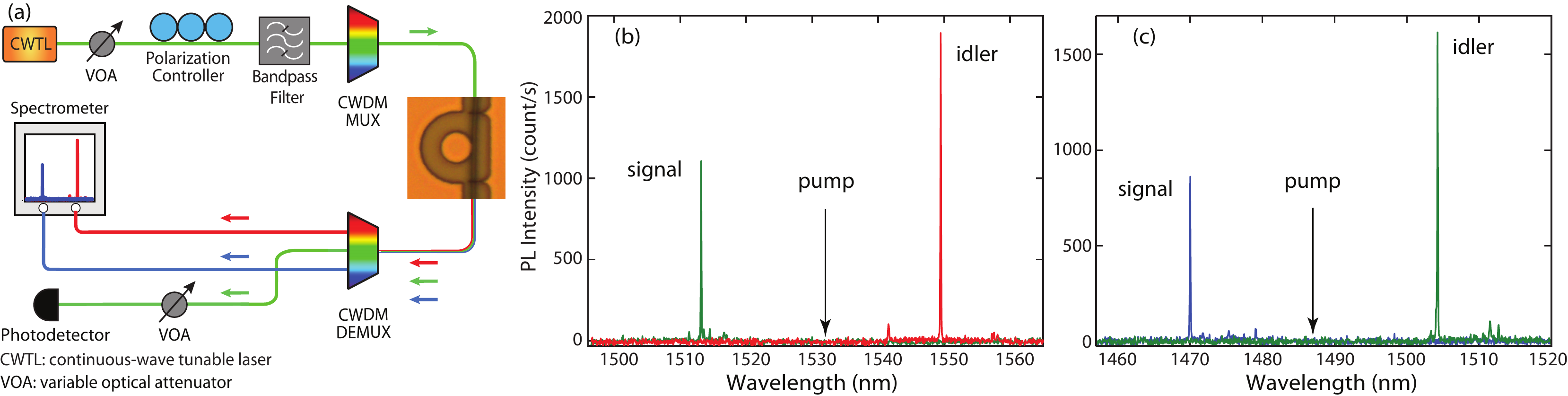}}
\end{figure*}

The demonstrated approach for precise dispersion engineering exhibits great potential for broad applications of nonlinear parametric processes. To show the power of the demonstrated technique, we apply it to achieve highly efficient photon-pair generation with precisely engineerable emission wavelengths, based on cavity-enhanced spontaneous four-wave mixing (SFWM). The experimental setup for measuring SFWM in our microdisk resonators is shown in Fig.~\ref{PLSetup}(a). The pump laser passes through a bandpass filter and a coarse wavelength-division multiplexer (CWDM MUX) to cut the laser noises, and is then coupled into the device via the tapered fiber. The pump wavelength is launched at the cavity resonance for TM${_2}$ mode where the ZDWL is located. The device output which consists of the optical pump and the generated signal and idler is separated into individual channel by a CWDM demultiplexer (DEMUX). The photoluminescence (PL) spectra of the signal and idler are recorded at each transmission port of the DEMUX for easy suppression of the pump wave. Fig.~\ref{PLSetup}(b) and (c) show the SFWM spectra for the microdisk without oxidation (pumping at 1532.2 nm) and that with an oxidation thickness of 12 nm (pumping at 1486.9 nm), respectively. They clearly show that, by precisely tailoring the ZDWL via thermal oxidation, a flexible selection of photon-pair emission wavelengths can be achieved. The spectrum of each emitted photon mode is so sharp that it is beyond the resolution of our spectrometer (${\sim}$0.135 nm), implying the high coherence of generated photons. The amplitude difference between the signal and idler is primarily due to different external couplings of cavity modes to the tapered fiber. When the pump mode is critically coupled to the cavity, the signal at shorter wavelength is under-coupled while the idler at longer wavelength is over-coupled, resulting in a higher photon extraction efficiency of the idler for both cases.

In summary, we have proposed and demonstrated a convenient and powerful CMOS-compatible technique for the precise dispersion management via silicon thermal oxidation. The demonstrated dispersion engineering of high-Q silicon microdisk resonators shows that thermal oxidation not only provides precise control of the ZDWL to achieve the phase-matching for the parametric process, but also reduces the sidewall roughness to improve the optical quality. Although we use the microdisk as an example, the demonstrated technique can readily be applied for any type of silicon waveguides/microresonators, such as microrings, photonic crystals, etc. Such a highly accurate dispersion management technique immediately allows us to achieve SFWM with precisely engineerable emission wavelengths, which shows great potential for applications in integrated silicon quantum photonics.

The work of W. Jiang, J. Zhang, and Q. Lin was supported by NSF and DARPA.
The work of N. Usechak was supported by AFRL's section 219 CRDF funding and the Air Force Office of Scientific Research under Grant 12RY09COR.

This work was performed in part at the Cornell NanoScale Facility, a member of the National Nanotechnology Infrastructure Network, which is supported by the National Science Foundation (Grant ECS-0335765).

\nocite{*}


\end{document}